\documentclass[a4paper]{article}

\usepackage{newtxtext} 
\usepackage{newtxmath}
\usepackage[T1]{fontenc}
\usepackage{multicol}
\usepackage{titlesec}
\usepackage{enumerate}
\usepackage{enumitem}
\usepackage{booktabs}
\usepackage{array}
\usepackage{hyperref}
\usepackage{fancyhdr}
\usepackage{tabularx}
\usepackage{lipsum}
\usepackage{graphicx}
\usepackage{soul,color}
\usepackage{cite}

\setlist[itemize]{leftmargin=*}
\setlist{noitemsep}

\titleformat{\subsection}      
  {\itshape}                      
  {\thesubsection.}               
  {0.5em}                         
  {}                              
\titlespacing*{\subsection}{0pt}{1ex}{1em} 

\titleformat{\subsubsection}[runin]
  {\itshape}                      
  {\thesubsubsection.}            
  {0.5em}                         
  {}
\titlespacing*{\subsubsection}{0pt}{1.5ex plus .1ex minus .2ex}{1.5ex plus .2ex}

\usepackage[
    top=1in,
    bottom=1in,
    left=0.59in,
    right=0.59in,
    headheight=61pt, 
    headsep=0.2in, 
    footskip=0.49in
]{geometry}

\makeatletter
\def\ciredmaketitle{
    \vspace*{-1 cm} 
    
    \begin{center}
        {\fontsize{18pt}{22pt}\selectfont \bfseries \@title \par} 
        \vspace{0.3cm} 
        
        {\lineskip 0.5em 
        {\fontsize{12pt}{14pt}\selectfont \bfseries \itshape \@author \par}
        }
        
        \vspace{-.9 cm} 
        
        \@date \par
    \end{center}
    \par
    \vspace*{0.5cm} 
}
\makeatother

\begin{document}

\title{Importance of Aggregated DER Installed Capacity in\\Distribution Networks}
\author{Alexandre M. V. Gouveia$^{1,2,3*}$, Md. Umar Hashmi$^{1,2}$, Reinhilde D'hulst$^{2,3,4}$, Dirk Van Hertem$^{1,2}$} 
\date{} 


\ciredmaketitle 
 \begin{center}

    $^1$KU Leuven, Leuven, Belgium \\
    $^2$EnergyVille, Genk, Belgium \\
    $^3$VITO, Mol, Belgium \\
    $^4$Orion Grid Technologies, Genk, Belgium \\
    $^*$alexandremiguel.venanciogouveia@kuleuven.be%

 \end{center}
 \begin{center}
  \end{center}
\begin{multicols}{2}

\section*{Abstract}
The increasing penetration of Distributed Energy Resources (DERs), particularly electric vehicles, heat pumps, and photovoltaic systems, is fundamentally changing power flows in Low-Voltage (LV) distribution networks. Despite this transition, Distribution System Operators (DSOs) often lack reliable and up-to-date knowledge of the DER capacity connected downstream of LV substations. Limited observability, incomplete topology information, and restricted access to customer-level data make it difficult to maintain accurate DER registries, creating uncertainty in both operational and planning processes. This paper presents aggregated DER installed capacity, estimated at LV aggregation points, as a practical and scalable approach to improving DER awareness without requiring customer‑level monitoring. We define the problem of estimating DER installed capacities from commonly available substation and feeder measurements. By linking these estimates to operational and planning needs, we discuss how knowledge of aggregated DER installed capacity enhances DER‑aware forecasting, congestion management, flexibility quantification, hosting capacity assessment, and monitoring of DER adoption.

\section{Introduction}

In many feeders, Distributed Energy Resources (DER)-related demand and generation already represent a substantial share of total energy flows, with Electric Vehicle (EV) charging, Heat Pump (HP) consumption and Photovoltaic (PV) generation fundamentally changing Low Voltage (LV) load profiles \cite{heider2024integration}. Unlocking the full potential of DERs for flexibility, planning, and operational purposes requires comprehensive, accurate, and up-to-date DER metadata. DER metadata encompasses information such as installed capacity (e.g. measured in kWp for PV, kVA for inverter interfacing assets, kW for EV charger and HPs), operating limits for active and/or reactive power or energy or ramp rate, geolocation, and data exchange and control capabilities. This paper will focus on DER installed capacity data, and this term will be used interchangeably with metadata throughout the rest of the paper.

At the LV level, Distribution System Operators (DSOs) typically lack full observability of the network and connected assets \cite{sanchez2017observability,barbato2018lessons}. LV topology models are often incomplete, feeder connectivity may be uncertain, and phase allocation is frequently unknown. Even with ongoing smart meter rollouts, measurements are commonly available only in aggregated form at secondary substations or feeders. Under these conditions, estimating aggregated DER installed capacity data provides a practical way to transform available measurements into actionable system knowledge.

Distribution network actors, such as DSOs, flexibility providers, aggregators, and market operators, significantly benefit from improved DER information. This paper addresses two key questions:

(i) How can DER installed capacity data be estimated using data already available to DSOs?

(ii) How can aggregated DER installed capacity data improve planning and operational decision-making?

This paper shows that regularly updated DER installed capacity metadata is essential for realistic LV network evaluation. Existing planning studies often assume greenfield or maximum penetration scenarios, which are unrealistic if they do not consider actual current DER penetration levels. Moreover, DSOs, often regulated monopolies, have limited transfer learning across regions, making local, measurement‑based DER visibility crucial. This work aims to show how accurate, aggregated DER metadata improves key operational and planning tasks—including forecasting, congestion management (CM), flexibility quantification (FQ), and hosting capacity assessment (HCA)—and to encourage stakeholders to keep this metadata regularly and accurately updated.

\section{Current State of DER Metadata}

DER metadata is typically collected through connection applications, installation certificates, and customer registries. While these mechanisms provide useful information at commissioning, metadata quality often degrades over time due to device replacement, ownership changes, undocumented upgrades, and inconsistent reporting requirements across DER \cite{dippenaar2025understanding,Horowitz2019DERInterconnection}.

In LV networks, several structural limitations prevent accurate DER tracking. Secondary substations often have limited monitoring, feeder models may be incomplete, and phase connectivity is rarely verified after installation. Privacy constraints and regulatory frameworks can further restrict access to customer-level DER information.

As a result, DSOs frequently operate with incomplete knowledge of the DER fleet installed at the LV level \cite{hickey2025controllability}. Planning tools (e.g. HCA) therefore rely on simplified assumptions or conservative estimates. Similarly, tools for CM or FQ operate suboptimally due to uncertainty regarding available controllable resources.

Aggregated DER metadata aligns with the spatial resolution of many DSO measurements and decisions. Instead of reconstructing individual installations, the objective becomes estimating the total installed DER capacity and technology composition connected downstream of a measurement point.

\section{Estimation of DER Metadata}


Aggregated metadata is particularly suited to LV operation and planning, where decisions such as transformer loading assessment, voltage management, and HCA are performed at feeder or substation level. Installed DER capacities can be estimated using measurements already available to DSOs, such as substation power flows, feeder measurements, or aggregated smart-meter data. The estimation relies on the fact that different DER technologies exhibit characteristic temporal signatures in aggregated load or generation measurements, often correlated with exogenous data (e.g., weather).

This problem is closely related to non-intrusive load monitoring (NILM) and load disaggregation, where individual appliance or technology consumption is inferred from aggregated measurements. This usually relates to Behind-the-Meter (BTM) detection of devices at the residential level, with high-frequency data. NILM methods based on feature extraction, optimisation, and machine learning have been studied in \cite{NILMRev}.

The installed capacity detection problem can be formulated as follows. Let $P_{\text{Net}}$ be the net load measured at a particular aggregation point (e.g., feeder head or substation)
\begin{align}
    &P_{\text{Net}} = P_{\text{Load}} - P_{\text{Gen}} \\
    &P_{\text{Load}} = P_{\text{NonDER}} + f(C_{\text{V1G}},C_{\text{HP}},C_{\text{BESS}}, \alpha)\\ 
    &P_{\text{Gen}} = g(C_{\text{PV}}, C_{\text{V2G}},C_{\text{BESS}}, \beta)
\end{align}
The aggregated consumption, $P_{\text{Load}}$, contains a traditional load component, $P_{\text{NonDER}}$, and a DER load component, modelled as a function $f(\cdot)$ of V1G, HP and BESS installed capacities ($C_{\text{V1G}}$, $C_{\text{HP}}$ and $C_{\text{BESS}}$) and relevant exogenous parameters ($\alpha$). Aggregated generation, $P_{\text{Gen}}$, can similarly be modelled as a function, $g(\cdot)$, of relevant installed capacities ($C_{\text{PV}}$, $C_{\text{V2G}}$ and $C_{\text{BESS}}$) and external parameters ($\beta$).

The goal of DER installed capacity estimation is to determine the relevant $C$ values, through $P_{\text{Net}}$, $\alpha$ and $\beta$ measurements, and a model of $f$ and $g$. Functions $f$ and $g$ can be modelled explicitly, for Model-Based approaches \cite{toro2023net,AYAD2025111059,gouveia2026pvcapacity}, or implicitly for Data-Driven approaches \cite{gouveia2026pvcapacity,MEDPOWER24}.

Several works have specifically addressed DER-relevant technologies, aggregated for multiple consumers. For example, event detection for EV charging and PV generation using aggregated measurements has been demonstrated using data-driven approaches \cite{NunesCIRED,jaramillo2023distributed}. Beyond event detection, recent work has explored direct estimation of aggregated DER capacity or load. Using representative meter selection and deep learning models, \cite{HUO2025135162} estimates regional EV charging demand. In \cite{MEDPOWER24}, a supervised ML model is trained to estimate aggregated PV capacity, using daily net load statistics. Similarly, HP load modelling from aggregated net load and annual heating demand has been investigated using linear temperature dependencies, typical HP data and HP ownership statistics \cite{AYAD2025111059}.

These approaches demonstrate that aggregated DER behaviour can be inferred without customer-level monitoring infrastructure. Because aggregated DER metadata evolves slowly compared to power measurements, periodic estimation using historical data is sufficient for planning and operational decision support.

\section{Use Cases of DER Metadata}

\begin{figure*}[t]
    \centering
    \includegraphics[width=0.74\linewidth]{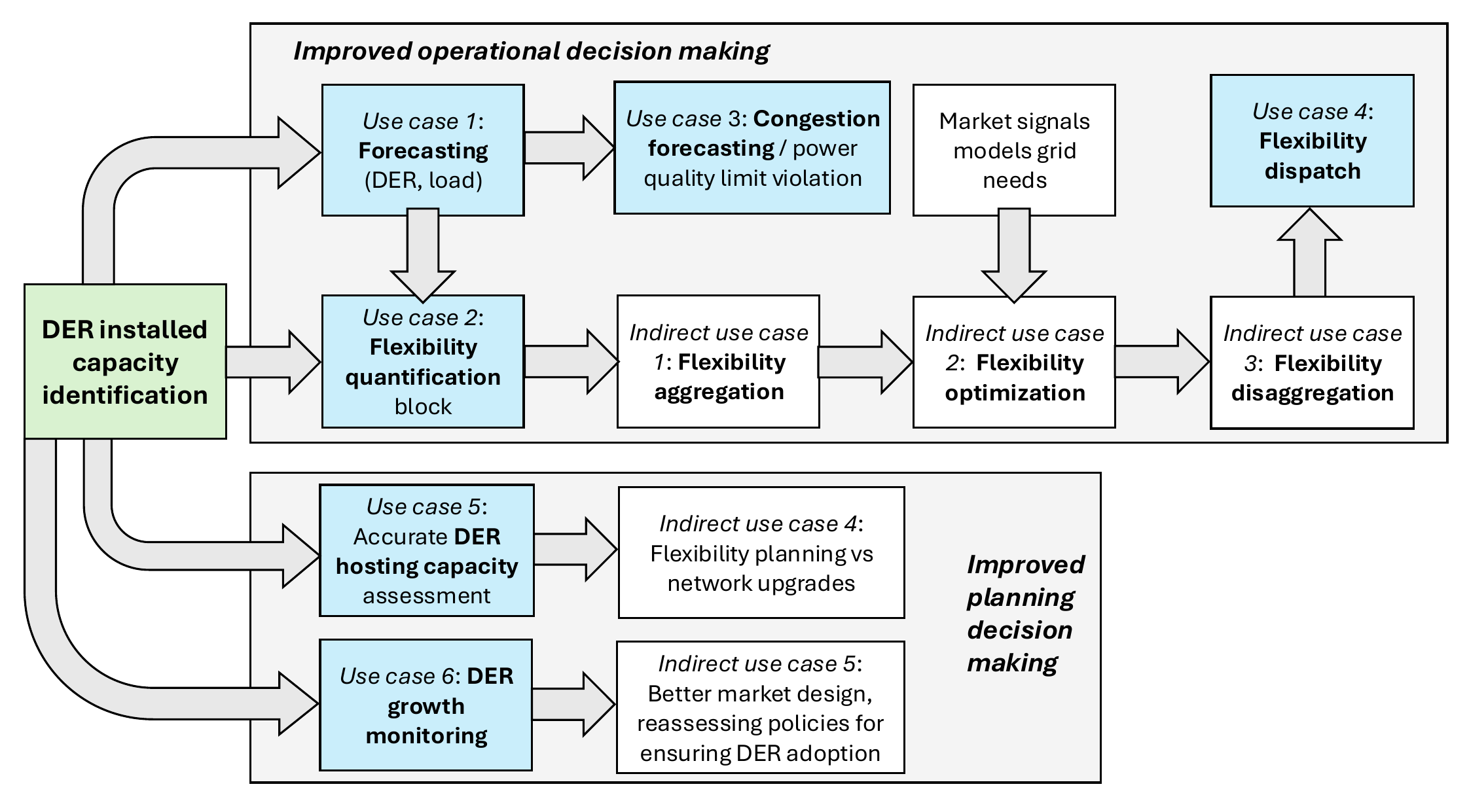}
    \vspace{-5pt}
    \caption{Flowchart of Use Cases for aggregated DER metadata.}
    \label{fig:use-cases}
\end{figure*}

Aggregated DER installed-capacity identification enables a chain of operational and planning applications for DSOs, as illustrated in Fig.~\ref{fig:use-cases}.

\subsection{DSO Operation}
DSO operation involves maintaining secure and reliable power delivery in distribution networks by ensuring voltage limits, thermal loading, and power-quality constraints are respected. In LV networks, operational awareness is typically limited to measurements at secondary substations, requiring DSOs to rely on forecasting, congestion anticipation, and coordination with flexibility providers rather than direct monitoring of individual assets. As DER penetration increases, understanding the composition of downstream demand and generation becomes increasingly important for anticipating peak loading, managing voltage conditions, and activating flexibility in a timely and effective manner.

\subsubsection{Forecasting}

DER-aware forecasting benefits from knowledge of installed DER capacity and technology mix. Instead of modelling load as a single aggregate signal, agnostic to embedded DERs, DER behaviours can be modelled explicitly. This can be done by using DER installed capacity as a feature or parameter of the net load forecasting model \cite{toro2023net}, by forecasting traditional load and DER profiles separately and combining them into a net load forecast \cite{thepprom,stratman2023}, and by scaling per-unit forecasts with estimated DER installed capacity \cite{nikzad2024estimating}. 

In \cite{toro2023net}, a modified bissection method is proposed for estimating aggregated installed PV capacity. The estimations are then fed to an ARX model for net load forecasting at LV substation level. Results show that the inclusion of PV capacity information led to an 11\% reduction in forecasting error, compared to estimating net load directly. Both \cite{thepprom} and \cite{stratman2023} disaggregate BTM PV generation and load to train independent neural networks for load and PV generation forecasting. These independent forecasts, and subsequent addition into net load forecasts, are compared with direct net load forecasts, using identically structured neural networks. Both works applied their method to data from Austin, Texas, from the Pecan Street Dataset \cite{pecan}. \cite{thepprom} cites an average 13\% reduction in RMSE by using independent forecasts across three different deep learning models, while \cite{stratman2023} shows a 16\% reduction in RMSE by using independent Long Short Term Memory (LSTM) networks, as opposed to one LSTM to directly forecast net load. In \cite{nikzad2024estimating}, a combined Convolutional Neural Network and LSTM (or CNN-LSTM) is used to forecast per-unit generation from PV and Wind, as well as consumption. An iterative process to estimate the corresponding installed capacities is also proposed, whose outputs are used to scale the per-unit forecasts. The resulting net load forecast is compared with a direct net load forecast using no estimated DER knowledge. The DER-aware CNN-LSTM showed consistently higher forecasting accuracy \textit{versus} the DER-agnostic one (93.70\% vs. 89.32\%).

Having improved accuracy in day-ahead forecasting can lead to more informed preemptive decisions on an operational timescale, as shown in Fig.~\ref{fig:use-cases}, and described further in this paper.

\subsubsection{Congestion Management}

Uncontrolled DERs tend to have synchronised behaviours linked to exogenous variables: HP demand is correlated with temperature, PV generation with irradiance, and EV charging with daily human activity. Higher concentrations of any given type of DER can then lead to higher peak currents and under/overvoltage risks due to their profiles changing in tandem \cite{vo2017}. Therefore, being aware of the installed capacity of these DERs can inform DSOs of congestion risks downstream of specific aggregation points, especially in response to certain events (e.g., extreme cold weather leading to peak HP consumption).

Preventive CM relies on accurate short-term forecasts \cite{vo2017,bernecker2025quantifying,shen2018comprehensive}. Specifically, having more accurate nodal load forecasts led to a 97\% reduction in CM costs \cite{bernecker2025quantifying}, while reliable net load time series are used for dynamic tariff and network reconfiguration optimisation \cite{vo2017,shen2018comprehensive}.

Furthermore, knowing the penetration rates of DERs in specific locations provides DSOs with a map of resources that can be used to manage network congestion.\cite{pantovs2020market} highlights the importance of locational information of assets participating in market-based CM, proposing location-related block orders for aggregators with multiple connection points.

\subsubsection{Flexibility Quantification \& Dispatch}

Flexibility procurement and activation depend on understanding how much controllable demand or generation is available in a network area. Aggregated DER metadata provides an estimate of installed flexible capacity, such as EV charging infrastructure or electrically driven heating systems. This enables DSOs or aggregators to estimate flexibility envelopes and feasibility regions \cite{nossair,ULBIG2015155}. Furthermore, DER installed capacity can be used as input for more detailed FQ methods, used as either boundary conditions or for baseline computation \cite{LIND2024101688,muller}. The installed capacity of the DER was shown to be the most sensitive parameter in FQ, especially in the case of EVs \cite{heulens2026flexibility}.

Having a per-DER day-ahead forecasting of load and/or generation, enhanced by having aggregated installed capacity data, also enables a more accurate and time-dependent quantification of potential day-ahead flexibility. With knowledge of the distribution of flexible assets connected to MV/LV transformers, flexibility activations can be done in a more targeted manner, as most flexibility needs at the distribution level arise to solve locational issues (e.g., network congestions). 

For flexibility requests for system balancing purposes, a flexibility potential map could also facilitate TSO/DSO/aggregator coordination. DSOs could more easily allocate and validate flexibility requests coming from the transmission side, sending this information to aggregators, who would activate the appropriate assets or send the appropriate market signals \cite{lind2019}.

\subsection{DSO Planning}
Distribution network planning focuses on ensuring the grid can accommodate current and future demand and generation while respecting technical limits and maintaining reliability. Planning tasks include asset sizing/placement, reinforcement decisions, HCA, and long-term investment prioritisation. In LV networks with increasing DER penetration, planning decisions depend on assumptions regarding existing and future DER capacity. In the absence of reliable knowledge about connected DERs, planners often rely on simplified or conservative scenarios, which can lead to either overinvestment or underestimated constraint risks.

\subsubsection{Hosting Capacity Assessment:}
HCA determines the maximum additional DER capacity that can be connected to a distribution network without violating thermal, voltage, or protection constraints. In practice, many HCA studies adopt simplified or “\textit{greenfield}” assumptions, where either no DERs or stylised penetration levels are assumed at the start of the analysis \cite{hashmi2025derhostingcapacitydistribution}. While such approaches are computationally convenient, they may misrepresent actual network conditions, particularly in LV grids with heterogeneous and rapidly evolving DER penetration.

Incorporating aggregated DER installed capacity information at the substation level enables metadata-aware HCA, where existing DERs are explicitly represented in the network model. HC is fundamentally a function of the current operating point and available network headroom. Existing PV generation reduces voltage rise margins during high-irradiance periods, while existing electrified heating or EV charging increases thermal loading during peak demand. Conversely, flexible loads can increase effective hosting capacity when coordinated control is assumed. Explicitly modelling installed DER capacity, therefore shifts HCA from a hypothetical to a condition-based evaluation.

The methodological foundations of HCA are typically power-flow-based constraint analyses, where voltage magnitude, thermal limits, and sometimes protection constraints are evaluated under worst-case scenarios (e.g., iterative or optimisation-based HCA formulations). Reviews of HCA methodologies highlight the sensitivity of results to assumptions regarding DER penetration, load composition, and network operating conditions. Deterministic, stochastic, and time-series HCA methods all rely on assumptions about DER capacity, allocation, and diversity, while probabilistic approaches explicitly model uncertainty in DER configuration and load behaviour \cite{MULENGA2020105445}. Consequently, inaccurate knowledge of installed DER capacity can lead to systematic over- or underestimation of hosting capacity.

\subsubsection{DER Growth Monitoring}
Monitoring the spatial evolution of DER adoption is becoming increasingly important for distribution network planning and operation. Electrification trends, such as the uptake of EVs, HPs, and PV systems, are rarely uniform across a service territory. Instead, DER adoption tends to cluster geographically due to factors such as housing characteristics, infrastructure availability, and socioeconomic conditions \cite{KONZEN2024103473}. Maintaining visibility of these uneven adoption patterns is difficult when DER registries are incomplete or not regularly updated.

Aggregated DER metadata anchored at secondary or primary substations provides a practical mechanism for geographically resolved DER growth monitoring. By periodically estimating installed DER capacity downstream of substations using available measurements, DSOs can track technology-specific penetration over time without relying on customer-level data. This enables the construction of spatial indicators of electrification across the LV network, allowing DSOs to identify feeders or substations where DER adoption is accelerating more rapidly than elsewhere.

Such monitoring makes it possible to detect uneven DER adoption across geographic areas, which may reflect differences in building stock, income levels, access to incentives, or infrastructure readiness. Identifying these disparities can support coordination between DSOs, regulators, and policymakers by providing evidence for locational or targeted incentive schemes aimed at promoting more equitable DER adoption \cite{Carley2020}. For example, areas with low electrification uptake despite suitable network capacity could be prioritised for awareness campaigns, financial incentives, or infrastructure support, while highly electrified areas may require complementary flexibility programs.

From a network perspective, geographically resolved DER growth indicators also support traditional planning decisions. Rapid increases in DER penetration at specific substations can signal the need for earlier reinforcement actions, such as transformer upgrades, feeder reinforcement, or voltage-control improvements. Conversely, areas with slower adoption may allow reinforcement investments to be deferred. In this way, DER growth monitoring complements conventional load-growth analysis by providing visibility into changes in load composition and generation patterns, rather than only total demand evolution.

Because installed DER capacity evolves on timescales of months to years, periodic estimation using historical measurements is sufficient to maintain an updated view of DER penetration. Aggregated DER metadata can therefore act as a continuously updated electrification indicator for LV networks, enabling DSOs to link DER adoption trends to both infrastructure planning and broader policy objectives related to equitable energy transition outcomes.


\section{Conclusions}
Increasing distributed energy resources (DER) penetration furthers the need for technology-aware operational and planning decisions. The lack of visibility on LV-connected DERs hinders DSO decision-making. Aggregated DER installed capacity estimation provides a practical mechanism to improve knowledge of downstream DER composition using existing measurements. This paper discussed how aggregated DER metadata can be inferred from distribution network measurements and analysed how such information enhances forecasting, congestion management, flexibility quantification, hosting capacity assessment, and DER growth monitoring. By linking measurement-based DER awareness with both operational processes and planning workflows, aggregated DER metadata provides a scalable approach to improving LV network observability without requiring customer-level monitoring. This work highlights the importance of DER metadata for distribution system stakeholders. Future work should focus on further improving methods for estimating DER installed capacity, as well as testing the integration of metadata-aware methods into operational and planning use cases, quantifying the impact of DER metadata knowledge.


\section*{Acknowledgments}
This work was partially funded by the KU Leuven funded C2-project FlexIQ (C2M/24/028).

\small

\bibliographystyle{IEEEtran}
\bibliography{reference.bib}

\begin{thebibliography}{10}
\providecommand{\url}[1]{#1}
\csname url@samestyle\endcsname
\providecommand{\newblock}{\relax}
\providecommand{\bibinfo}[2]{#2}
\providecommand{\BIBentrySTDinterwordspacing}{\spaceskip=0pt\relax}
\providecommand{\BIBentryALTinterwordstretchfactor}{4}
\providecommand{\BIBentryALTinterwordspacing}{\spaceskip=\fontdimen2\font plus
\BIBentryALTinterwordstretchfactor\fontdimen3\font minus \fontdimen4\font\relax}
\providecommand{\BIBforeignlanguage}[2]{{%
\expandafter\ifx\csname l@#1\endcsname\relax
\typeout{** WARNING: IEEEtran.bst: No hyphenation pattern has been}%
\typeout{** loaded for the language `#1'. Using the pattern for}%
\typeout{** the default language instead.}%
\else
\language=\csname l@#1\endcsname
\fi
#2}}
\providecommand{\BIBdecl}{\relax}
\BIBdecl

\bibitem{heider2024integration}
A.~Heider, K.~Helfenbein, B.~Schachler, T.~R{\"o}pcke, and G.~Hug, ``On the integration of electric vehicles into german distribution grids through smart charging,'' \emph{IEEE Transactions on Industry Applications}, vol.~61, no.~2, pp. 2001--2010, 2024.

\bibitem{sanchez2017observability}
R.~Sanchez, F.~Iov, M.~Kemal, M.~Stefan, and R.~Olsen, ``Observability of low voltage grids: Actual dsos challenges and research questions,'' in \emph{2017 52nd International Universities Power Engineering Conference (UPEC)}.\hskip 1em plus 0.5em minus 0.4em\relax IEEE, 2017, pp. 1--6.

\bibitem{barbato2018lessons}
A.~Barbato \emph{et~al.}, ``Lessons learnt from real-time monitoring of the low voltage distribution network,'' \emph{Sustainable Energy, Grids and Networks}, vol.~15, pp. 76--85, 2018.

\bibitem{dippenaar2025understanding}
J.~A. Dippenaar and B.~Bekker, ``Understanding interconnection rule non-compliance: Lessons from south africa's surge in unauthorised distributed energy resources,'' \emph{Energy for Sustainable Development}, vol.~85, p. 101661, 2025.

\bibitem{Horowitz2019DERInterconnection}
K.~Horowitz \emph{et~al.}, ``{An Overview of Distributed Energy Resource (DER) Interconnection: Current Practices and Emerging Solutions},'' National Renewable Energy Laboratory (NREL), Golden, CO, USA, Tech. Rep. NREL/TP-6A20-72102, 2019.

\bibitem{hickey2025controllability}
M.~Hickey, C.~de~Brito, P.~S. Alves, F.~O’Donnell, and A.~Ahmadi, ``{Controllability, visibility, forecasting and modelling strategies to increase distributed energy resources integration in Ireland},'' in \emph{28th International Conference and Exhibition on Electricity Distribution (CIRED 2025)}, vol. 2025.\hskip 1em plus 0.5em minus 0.4em\relax IET, 2025, pp. 1037--1041.

\bibitem{NILMRev}
P.~A. Schirmer and I.~Mporas, ``Non-intrusive load monitoring: A review,'' \emph{IEEE Transactions on Smart Grid}, vol.~14, no.~1, pp. 769--784, 2023.

\bibitem{toro2023net}
M.~Toro-C{\'a}rdenas, I.~Moreira, H.~Morais, P.~M. Carvalho, and L.~A. Ferreira, ``Net load disaggregation at secondary substation level,'' \emph{Renewable Energy}, vol. 207, pp. 765--771, 2023.

\bibitem{AYAD2025111059}
A.~Ayad, S.~Wong, and V.~Delisle, ``Modeling of heat pumps load profiles for power systems integration,'' \emph{Electric Power Systems Research}, vol. 238, p. 111059, 2025.

\bibitem{gouveia2026pvcapacity}
A.~M.~V. Gouveia, M.~U. Hashmi, R.~D'hulst, and D.~Van~Hertem, ``{Installed PV Capacity Detection on LV Substations: Comparison of Data-Driven and Model-Based Methods},'' 2026, available at SSRN: \url{https://ssrn.com/abstract=5717615}.

\bibitem{MEDPOWER24}
A.~M.~V. Gouveia, R.~D’hulst, M.~U. Hashmi, and D.~Van~Hertem, ``Data-driven photovoltaic installed capacity detection at low voltage substation level,'' in \emph{14th Mediterranean Conference on Power Generation Transmission, Distribution and Energy Conversion (MEDPOWER 2024)}, 2024, pp. 678--683.

\bibitem{NunesCIRED}
M.~L. Nunes, J.~F.~P. Fernandes, J.~Oliveira, and P.~J. Costa~Branco, ``Power factor signature analysis for disaggregation of ev charging loads from aggregated power,'' in \emph{CIRED Porto Workshop 2022: E-mobility and power distribution systems}, vol. 2022, 2022, pp. 23--27.

\bibitem{jaramillo2023distributed}
A.~F.~M. Jaramillo \emph{et~al.}, ``Distributed energy resources electric profile identification in low voltage networks using supervised machine learning techniques,'' \emph{IEEE Access}, vol.~11, pp. 19\,469--19\,486, 2023.

\bibitem{HUO2025135162}
Y.~Huo \emph{et~al.}, ``Real-time estimation of aggregated electric vehicle charging load based on representative meter data,'' \emph{Energy}, vol. 321, p. 135162, 2025.

\bibitem{thepprom}
C.~Thepprom, N.~Nupairoj, and P.~Vateekul, ``A deep learning framework for net load forecasting with unsupervised behind-the-meter disaggregated data,'' \emph{IEEE Access}, vol.~12, pp. 94\,958--94\,971, 2024.

\bibitem{stratman2023}
A.~Stratman, T.~Hong, M.~Yi, and D.~Zhao, ``Net load forecasting with disaggregated behind-the-meter pv generation,'' \emph{IEEE Transactions on Industry Applications}, vol.~59, no.~5, pp. 5341--5351, 2023.

\bibitem{nikzad2024estimating}
A.~R. Nikzad, A.~A. Mohamed, B.~Venkatesh, and J.~Penaranda, ``Estimating aggregate capacity of connected ders and forecasting feeder power flow with limited data availability,'' \emph{IEEE Open Access Journal of Power and Energy}, vol.~11, pp. 266--279, 2024.

\bibitem{pecan}
{Pecan Street Inc.}, ``Dataport,'' \url{https://www.pecanstreet.org/}, 2021.

\bibitem{vo2017}
T.~H. Vo, A.~N. M.~M. Haque, P.~H. Nguyen, I.~G. Kamphuis, M.~Eijgelaar, and I.~Bouwman, ``A study of congestion management in smart distribution networks based on demand flexibility,'' in \emph{2017 IEEE Manchester PowerTech}, 2017, pp. 1--6.

\bibitem{bernecker2025quantifying}
M.~Bernecker, M.~Gebhardt, S.~B. Amor, M.~Wolter, and F.~M{\"u}sgens, ``Quantifying the impact of load forecasting accuracy on congestion management in distribution grids,'' \emph{International Journal of Electrical Power \& Energy Systems}, vol. 168, p. 110713, 2025.

\bibitem{shen2018comprehensive}
F.~Shen, S.~Huang, Q.~Wu, S.~Repo, Y.~Xu, and J.~{\O}stergaard, ``Comprehensive congestion management for distribution networks based on dynamic tariff, reconfiguration, and re-profiling product,'' \emph{IEEE Transactions on Smart Grid}, vol.~10, no.~5, pp. 4795--4805, 2018.

\bibitem{pantovs2020market}
M.~Panto{\v{s}}, ``Market-based congestion management in electric power systems with exploitation of aggregators,'' \emph{International Journal of Electrical Power \& Energy Systems}, vol. 121, p. 106101, 2020.

\bibitem{nossair}
H.~Nosair and F.~Bouffard, ``Flexibility envelopes for power system operational planning,'' \emph{IEEE Transactions on Sustainable Energy}, vol.~6, no.~3, pp. 800--809, 2015.

\bibitem{ULBIG2015155}
A.~Ulbig and G.~Andersson, ``Analyzing operational flexibility of electric power systems,'' \emph{International Journal of Electrical Power \& Energy Systems}, vol.~72, pp. 155--164, 2015, the Special Issue for 18th Power Systems Computation Conference.

\bibitem{LIND2024101688}
L.~Lind, J.~P. Chaves-Ávila, O.~Valarezo, A.~Sanjab, and L.~Olmos, ``Baseline methods for distributed flexibility in power systems considering resource, market, and product characteristics,'' \emph{Utilities Policy}, vol.~86, p. 101688, 2024.

\bibitem{muller}
F.~L. Müller, J.~Szabó, O.~Sundström, and J.~Lygeros, ``Aggregation and disaggregation of energetic flexibility from distributed energy resources,'' \emph{IEEE Transactions on Smart Grid}, vol.~10, no.~2, pp. 1205--1214, 2019.

\bibitem{heulens2026flexibility}
L.~Heulens, M.~U. Hashmi, A.~M.~V. Gouveia, and D.~Van~Hertem, ``Flexibility quantification of electric vehicle, heat pump, and photovoltaic generation considering process, comfort constraints, and grid needs,'' 2026, unpublished manuscript.

\bibitem{lind2019}
L.~Lind, R.~Cossent, J.~P. Chaves-Ávila, and T.~Gómez San~Román, ``Transmission and distribution coordination in power systems with high shares of distributed energy resources providing balancing and congestion management services,'' \emph{WIREs Energy and Environment}, vol.~8, no.~6, p. e357, 2019.

\bibitem{hashmi2025derhostingcapacitydistribution}
M.~U. Hashmi, ``Der hosting capacity for distribution networks: definitions, attributes, use-cases and challenges,'' in \emph{IET Conference Proceedings CP922}, vol. 2025, no.~14.\hskip 1em plus 0.5em minus 0.4em\relax IET, 2025, pp. 541--546.

\bibitem{MULENGA2020105445}
E.~Mulenga, M.~H. Bollen, and N.~Etherden, ``A review of hosting capacity quantification methods for photovoltaics in low-voltage distribution grids,'' \emph{International Journal of Electrical Power \& Energy Systems}, vol. 115, p. 105445, 2020.

\bibitem{KONZEN2024103473}
G.~Konzen, R.~Best, and N.~J. {de Castro}, ``The energy injustice of household solar energy: A systematic review of distributional disparities in residential rooftop solar adoption,'' \emph{Energy Research \& Social Science}, vol. 111, p. 103473, 2024.

\bibitem{Carley2020}
S.~Carley and D.~M. Konisky, ``The justice and equity implications of the clean energy transition,'' \emph{Nat Energy}, vol.~5, pp. 569–--577, 2020.

\end{thebibliography}

\end{multicols}
\end{document}